\documentclass[floatfix,10pt,twocolumn,superscriptaddress,aps,prd,nofootinbib]{revtex4-2}
\usepackage{graphicx}
\usepackage{subfigure}
\usepackage{dcolumn}\usepackage{bm}
\usepackage[usenames]{color}
\usepackage{lineno}
\usepackage{amsmath}
\usepackage{amssymb}
\usepackage{etoolbox} 
\usepackage{url} 
\usepackage{placeins}
\usepackage{hyperref} 
\usepackage{enumitem}

\usepackage{multirow} 
\usepackage{orcidlink}

\newcommand*\linenomathpatch[1]{%
  \cspreto{#1}{\linenomath}%
  \cspreto{#1*}{\linenomath}%
  \csappto{end#1}{\endlinenomath}%
  \csappto{end#1*}{\endlinenomath}%
}
\linenomathpatch{equation}
\linenomathpatch{gather}
\linenomathpatch{multline}
\linenomathpatch{align}
\linenomathpatch{alignat}
\linenomathpatch{flalign}

\def\address{\@ifstar{\address@star}%
  {\@ifnextchar[{\address@optarg}{\address@noptarg}}}

\newcommand\etainv{ \eta^{(\prime)} \to invisible}
\newcommand\etap{ \eta^{(\prime)}}

\begin{document}

\title{EUROPEAN LABORATORY FOR PARTICLE PHYSICS\\
\vskip0.5cm
\hspace{-0.0cm}{\rightline{\rm  CERN-EP-2024-153}}
\vskip0.5cm
50 GeV $\pi^-$ in, nothing out: a sensitive  probe of invisible  $\eta$ and $\eta'$ decays with NA64h}

\author{Yu.~M.~Andreev\orcidlink{0000-0002-7397-9665}}
\affiliation{Authors affiliated with an institute covered by a cooperation agreement with CERN}
\author{A.~Antonov\orcidlink{0000-0003-1238-5158}}
\affiliation{INFN, Sezione di Genova, 16147 Genova, Italia}
\author{M.~A.~ Ayala Torres}
\affiliation{Center for Theoretical and Experimental Particle Physics, Facultad de Ciencias Exactas, Universidad Andres Bello, Fernandez Concha 700, Santiago, Chile}
\affiliation{Millennium Institute for Subatomic Physics at High-Energy Frontier (SAPHIR), Fernandez Concha 700, Santiago, Chile}
\author{D.~Banerjee\orcidlink{0000-0003-0531-1679}}
\affiliation{CERN, European Organization for Nuclear Research, CH-1211 Geneva, Switzerland}
\author{B.~Banto Oberhauser\orcidlink{0009-0006-4795-1008}}
\affiliation{ETH Z\"urich, Institute for Particle Physics and Astrophysics, CH-8093 Z\"urich, Switzerland}
\author{J.~Bernhard\orcidlink{0000-0001-9256-971X}}
\affiliation{CERN, European Organization for Nuclear Research, CH-1211 Geneva, Switzerland}
\author{P.~Bisio\orcidlink{/0009-0006-8677-7495}}
\affiliation{INFN, Sezione di Genova, 16147 Genova, Italia}
\affiliation{Universit\`a degli Studi di Genova, 16126 Genova, Italia}
\author{A.~Celentano\orcidlink{0000-0002-7104-2983}}
\affiliation{INFN, Sezione di Genova, 16147 Genova, Italia}
\author{N.~Charitonidis\orcidlink{0000-0001-9506-1022}}
\affiliation{CERN, European Organization for Nuclear Research, CH-1211 Geneva, Switzerland}
\author{D.~Cooke}
\affiliation{UCL Departement of Physics and Astronomy, University College London, Gower St. London WC1E 6BT, United Kingdom}
\author{P.~Crivelli\orcidlink{0000-0001-5430-9394}}
\thanks{Corresponding author}\email{paolo.crivelli@cern.ch}
\affiliation{ETH Z\"urich, Institute for Particle Physics and Astrophysics, CH-8093 Z\"urich, Switzerland}
\author{E.~Depero\orcidlink{0000-0003-2239-1746}}
\affiliation{ETH Z\"urich, Institute for Particle Physics and Astrophysics, CH-8093 Z\"urich, Switzerland}
\author{A.~V.~Dermenev\orcidlink{0000-0001-5619-376X}}
\affiliation{Authors affiliated with an institute covered by a cooperation agreement with CERN}
\author{S.~V.~Donskov\orcidlink{0000-0002-3988-7687}}
\affiliation{Authors affiliated with an institute covered by a cooperation agreement with CERN}
\author{R.~R.~Dusaev\orcidlink{0000-0002-6147-8038}}
\affiliation{Authors affiliated with an institute covered by a cooperation agreement with CERN}
\author{T.~Enik\orcidlink{0000-0002-2761-9730}}
\affiliation{Authors affiliated with an international laboratory covered by a cooperation agreement with CERN}
\author{V.~N.~Frolov}
\affiliation{Authors affiliated with an international laboratory covered by a cooperation agreement with CERN}
\author{S.~V.~Gertsenberger\orcidlink{0009-0006-1640-9443}}
\affiliation{Authors affiliated with an international laboratory covered by a cooperation agreement with CERN}
\author{S.~Girod}
\affiliation{CERN, European Organization for Nuclear Research, CH-1211 Geneva, Switzerland}
\author{S.~N.~Gninenko\orcidlink{0000-0001-6495-7619}}
\thanks{Corresponding author}\email{sergei.gninenko@cern.ch}
\affiliation{Authors affiliated with an institute covered by a cooperation agreement with CERN}
\affiliation{Millennium Institute for Subatomic Physics at High-Energy Frontier (SAPHIR), Fernandez Concha 700, Santiago, Chile}
\author{M.~H\"osgen}
\affiliation{Universit\"at Bonn, Helmholtz-Institut f\"ur Strahlen-und Kernphysik, 53115 Bonn, Germany}
\author{V.~A.~Kachanov\orcidlink{0000-0002-3062-010X}}
\affiliation{Authors affiliated with an institute covered by a cooperation agreement with CERN}
\author{Y.~Kambar\orcidlink{0009-0000-9185-2353}}
\affiliation{Authors affiliated with an international laboratory covered by a cooperation agreement with CERN}
\author{A.~E.~Karneyeu\orcidlink{0000-0001-9983-1004}}
\affiliation{Authors affiliated with an institute covered by a cooperation agreement with CERN}
\author{G.~D.~Kekelidze\orcidlink{0000-0002-5393-9199}}
\affiliation{Authors affiliated with an international laboratory covered by a cooperation agreement with CERN}
\author{B.~Ketzer\orcidlink{0000-0002-3493-3891}}
\affiliation{Universit\"at Bonn, Helmholtz-Institut f\"ur Strahlen-und Kernphysik, 53115 Bonn, Germany}
\author{D.~V.~Kirpichnikov\orcidlink{0000-0002-7177-077X}}
\affiliation{Authors affiliated with an institute covered by a cooperation agreement with CERN}
\author{V.~N.~Kolosov}
\affiliation{Authors affiliated with an institute covered by a cooperation agreement with CERN}
\author{V.~A.~Kramarenko\orcidlink{0000-0002-8625-5586}}
\affiliation{Authors affiliated with an institute covered by a cooperation agreement with CERN}
\affiliation{Authors affiliated with an international laboratory covered by a cooperation agreement with CERN}
\author{L.~V.~Kravchuk\orcidlink{0000-0001-8631-4200}}
\affiliation{Authors affiliated with an institute covered by a cooperation agreement with CERN}
\author{N.~V.~Krasnikov\orcidlink{0000-0002-8717-6492}}
\affiliation{Authors affiliated with an institute covered by a cooperation agreement with CERN}
\affiliation{Authors affiliated with an international laboratory covered by a cooperation agreement with CERN}
\author{S.~V.~Kuleshov\orcidlink{0000-0002-3065-326X}}
\thanks{Corresponding author}\email{serguei.koulechov@cern.ch}
\affiliation{Center for Theoretical and Experimental Particle Physics, Facultad de Ciencias Exactas, Universidad Andres Bello, Fernandez Concha 700, Santiago, Chile}
\affiliation{Millennium Institute for Subatomic Physics at High-Energy Frontier (SAPHIR), Fernandez Concha 700, Santiago, Chile}
\author{V.~E.~Lyubovitskij\orcidlink{0000-0001-7467-572X}}
\affiliation{Authors affiliated with an institute covered by a cooperation agreement with CERN}
\affiliation{Universidad T\'ecnica Federico Santa Mar\'ia and CCTVal, 2390123 Valpara\'iso, Chile}
\affiliation{Millennium Institute for Subatomic Physics at High-Energy Frontier (SAPHIR), Fernandez Concha 700, Santiago, Chile}
\author{A.~Marini\orcidlink{0000-0002-6778-2161}}
\affiliation{INFN, Sezione di Genova, 16147 Genova, Italia}
\author{L.~Marsicano\orcidlink{0000-0002-8931-7498}}
\affiliation{INFN, Sezione di Genova, 16147 Genova, Italia}
\author{V.~A.~Matveev\orcidlink{0000-0002-2745-5908}}
\affiliation{Authors affiliated with an international laboratory covered by a cooperation agreement with CERN}
\author{R.~Mena~Fredes}
\affiliation{Millennium Institute for Subatomic Physics at High-Energy Frontier (SAPHIR), Fernandez Concha 700, Santiago, Chile}
\affiliation{Universidad T\'ecnica Federico Santa Mar\'ia and CCTVal, 2390123 Valpara\'iso, Chile}
\author{R.~ G.~Mena~Yanssen}
\affiliation{Millennium Institute for Subatomic Physics at High-Energy Frontier (SAPHIR), Fernandez Concha 700, Santiago, Chile}
\affiliation{Universidad T\'ecnica Federico Santa Mar\'ia and CCTVal, 2390123 Valpara\'iso, Chile}
\author{L.~Molina Bueno\orcidlink{0000-0001-9720-9764}}
\affiliation{Instituto de Fisica Corpuscular (CSIC/UV), Carrer del Catedratic Jose Beltran Martinez, 2, 46980 Paterna, Valencia, Spain}
\author{M.~Mongillo\orcidlink{0009-0000-7331-4076}}
\affiliation{ETH Z\"urich, Institute for Particle Physics and Astrophysics, CH-8093 Z\"urich, Switzerland}
\author{D.~V.~Peshekhonov\orcidlink{0009-0008-9018-5884}}
\affiliation{Authors affiliated with an international laboratory covered by a cooperation agreement with CERN}
\author{V.~A.~Polyakov\orcidlink{0000-0001-5989-0990}}
\affiliation{Authors affiliated with an institute covered by a cooperation agreement with CERN}
\author{B.~Radics\orcidlink{0000-0002-8978-1725}}
\affiliation{York University, Toronto, Canada}
\author{K.~M.~Salamatin\orcidlink{0000-0001-6287-8685}}
\affiliation{Authors affiliated with an international laboratory covered by a cooperation agreement with CERN}
\author{V.~D.~Samoylenko}
\affiliation{Authors affiliated with an institute covered by a cooperation agreement with CERN}
\author{H.~Sieber\orcidlink{0000-0003-1476-4258}}
\affiliation{ETH Z\"urich, Institute for Particle Physics and Astrophysics, CH-8093 Z\"urich, Switzerland}
\author{D.~A.~Shchukin\orcidlink{0009-0007-5508-3615}}
\affiliation{Authors affiliated with an institute covered by a cooperation agreement with CERN}
\author{O.~Soto}
\affiliation{Departamento de Fisica, Facultad de Ciencias, Universidad de La Serena, Avenida Cisternas 1200, La Serena, Chile}
\affiliation{Millennium Institute for Subatomic Physics at High-Energy Frontier (SAPHIR), Fernandez Concha 700, Santiago, Chile}
\author{V.~O.~Tikhomirov\orcidlink{0000-0002-9634-0581}}
\affiliation{Authors affiliated with an institute covered by a cooperation agreement with CERN}
\author{I.~V.~Tlisova\orcidlink{0000-0003-1552-2015}}
\affiliation{Authors affiliated with an institute covered by a cooperation agreement with CERN}
\author{A.~N.~Toropin\orcidlink{0000-0002-2106-4041}}
\affiliation{Authors affiliated with an institute covered by a cooperation agreement with CERN}
\author{M.~Tuzi\orcidlink{0009-0000-6276-1401}}
\affiliation{Instituto de Fisica Corpuscular (CSIC/UV), Carrer del Catedratic Jose Beltran Martinez, 2, 46980 Paterna, Valencia, Spain}
\author{P.~V.~Volkov\orcidlink{0000-0002-7668-3691}}
\affiliation{Authors affiliated with an international laboratory covered by a cooperation agreement with CERN}
\author{V.~Yu.~Volkov\orcidlink{0009-0005-3500-5121}}
\thanks{Deceased}
\affiliation{Authors affiliated with an institute covered by a cooperation agreement with CERN}
\thanks{Desease}
\author{I.~V.~Voronchikhin\orcidlink{0000-0003-3037-636X}}
\affiliation{Authors affiliated with an institute covered by a cooperation agreement with CERN}
\author{J.~Zamora-Sa\'a\orcidlink{0000-0002-5030-7516}}
\affiliation{Center for Theoretical and Experimental Particle Physics, Facultad de Ciencias Exactas, Universidad Andres Bello, Fernandez Concha 700, Santiago, Chile}
\affiliation{Millennium Institute for Subatomic Physics at High-Energy Frontier (SAPHIR), Fernandez Concha 700, Santiago, Chile}
\author{A.~S.~Zhevlakov\orcidlink{0000-0002-7775-5917}}
\affiliation{Authors affiliated with an international laboratory covered by a cooperation agreement with CERN}
\collaboration{The NA64 Collaboration}\noaffiliation
\vskip 0.25cm

\date{\today}


\begin{abstract}
We present the first results from a proof-of-concept search for dark sectors via invisible decays of pseudoscalar $\eta$ and $\eta'$ mesons in the  NA64h experiment at the CERN SPS. Our novel technique uses the charge-exchange reaction of 50 GeV $\pi^-$ on nuclei of an active target as the source of neutral mesons. The $\eta, \eta' \to invisible$ events would exhibit themselves via a striking  signature - the complete disappearance of the incoming beam energy in the detector. No evidence for such events has been found  with  $2.9\times10^{9}$ pions on target accumulated during one day of data taking. This allows us to set a stringent limit on the branching ratio ${\rm Br}(\eta' \to invisible) < 2.1 \times 10^{-4}$ improving the current bound by a factor of $\simeq3$. We also set a limit on ${\rm Br}(\eta \to invisible) < 1.1 \times 10^{-4}$ comparable with the existing one. These results demonstrate the great potential of our approach and provide clear guidance on how to enhance and extend the sensitivity  for dark sector physics  from  future searches for invisible neutral meson decays. 

\end{abstract}

\pacs{14.80.-j, 12.60.-i, 13.20.Cz, 13.35.Hb}

\maketitle
Decays of the pseudoscalar neutral  mesons $(M^0)$, such as $\pi^0, \eta, \eta', K^0_S, K^0_L$,  provide a  unique opportunity to probe new physics beyond the Standard Model (SM) \cite{pdg}.  Searching for  their decay into invisible final states is particularly  advantageous because in the SM the branching fraction  of the  $M^0$  decay into a 
neutrino-antineutrino pair, 
${\rm Br}(M^0\to  \nu \overline{\nu})$,  is predicted to be   extremely small~\cite{marci}. Indeed, for massless neutrinos, this transition is forbidden kinematically by angular momentum conservation. 
For the case of massive neutrinos, one of them is forced to  have the  "wrong" helicity resulting  in the suppression of  ${\rm Br}(M^0\to  \nu \overline{\nu})$ by  a factor proportional to  the neutrino mass squared,   
$\sim m_\nu^2/m_{M^0}^2 \lesssim 10^{-16} $, 
for $m_\nu \lesssim 10$ eV and $m_{M^0} \simeq m_\eta \simeq  0.5$ GeV \cite{pdg}. In the SM the helicity suppression can be overcome for the four-neutrino final state, however,  in this case, ${\rm Br}(M^0 \to  \nu \overline{\nu} \nu \overline{\nu}) \lesssim 10^{-18}$~\cite{gao}. 
Thus, if the decay $M^0\to invisible$ is observed it would unambiguously signal the presence of new physics.
\par Various extensions of the SM  could significantly enhance the invisible  decay rate  of $\eta, \eta'$, and $K^0_S, K^0_L$ up to a measurable level, 
 for a  recent review see, e.g., Ref.~\cite{tulin}, and  Refs.~\cite{kam,gninenkoprd15,gk2015,gk2016,gabri,host,redtop},  respectively.
Some of  the scenarios consider dark sector physics, including light Dark Matter (DM),  with masses of the DM  particles ($\chi$) much below the electroweak scale, 
$m_\chi \ll \Lambda_{\rm EW} \simeq 100$  GeV, which has  a new interaction between the SM and DM transmitted by a scalar mediator \cite{fayet}. Such mediator naturally couples more strongly to DM than SM particles, hence it
would dominantly  decay invisibly  if kinematically allowed. The $\etap \to invisible$ process could occur via  the decay into pairs of mediators subsequently decaying 
to DM particles \cite{fayet}, or  from direct decay $\etap \to \chi \chi$  \cite{bob,darme}. 
An interesting case is when the (pseudo)scalar mediator is  leptophobic, i.e. transmits  interaction between the $\chi$ and light SM quarks,  and can accommodate the relic DM density, see, e.g., \cite{batell,ema}. 
Another attractive model considers  the $\etap \to invisible$  decay  into a pair of heavy neutrinos~\cite{li}.
\par Searching  for the $M^0\to invisible$ decay is challenging, as it requires a combination of  an intense source of  $M^0$s and a well-defined high-purity signature 
to tag their production. The most sensitive limit,   ${\rm Br}(M^0\to invisible) \lesssim 10^{-9}$,  is obtained for  $\pi^0$'s produced via  the $K^-\to \pi^- \pi^0$ decay  \cite{na62}. 
   Several previous searches for  $\eta,\eta' \to invisible$ decays  have been performed at $e^+e^-$ colliders by  the BES~\cite{bes}, CLEO~\cite{cleo}  and BESIII~\cite{besiii} experiments. However, the best upper limits  ${\rm Br}(\eta \to invisible) < 1.0\times 10^{-4}$ and ${\rm Br}(\eta' \to invisible) < 6.0\times 10^{-4}$ at 90\% confidence level (C.L.)  obtained by BESIII, are  still much less  stringent compared to the $\pi^0$ one. These bounds were obtained from a sample of $\sim 2.3\times 10^8~ J/\psi $ events collected during $\sim50$ days of running at the BEPCII 
 \cite{bepc}  by using $J/\psi \to \phi \eta, \phi \eta'; \phi\to K^+K^-$ decay chain as a source of tagged $\eta, \eta'$ \cite{besiii}.   
 \begin{figure}[tbh!]
\centering
\includegraphics[width=0.2\textwidth]{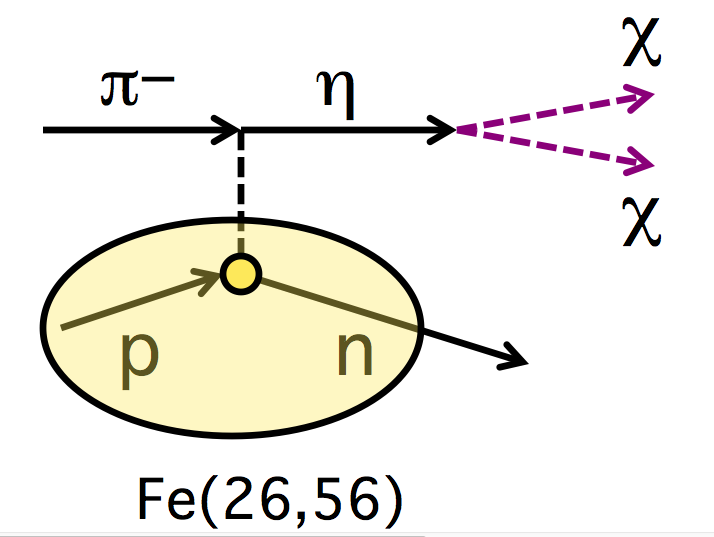}%
\caption{Diagrams illustrating the $\eta$ production in the quasi-elastic charge exchange  reaction of Eq.(\ref{eq:etaprod}).\label{fig:product}}
\end{figure} 
 \par In this Letter, we report the first results of the search for invisible $\eta, \eta'$ decays in the NA64h fixed-target active beam-dump experiment at the CERN Super Proton Synchrotron accelerator (SPS) \cite{ufn,pc} obtained from one day of data taking from a run with a hadron(h) beam. 
The method we chose for the search was proposed in Ref.~\cite{gninenkoprd15}. 
The source of the $\eta$ and $\eta'$ mesons is the quasi-elastic charge exchange  reaction of 50 GeV $\pi^-$'s on nuclei $A(Z)$  of an active target 
\begin{equation}
\pi^- + A(Z) \to  \etap + n + A(Z-1); \etap \to invisible
\label{eq:etaprod}
\end{equation}
as illustrated in Fig.\ref{fig:product}. 
\begin{figure}[tbh!!]
\includegraphics[width=.45\textwidth]{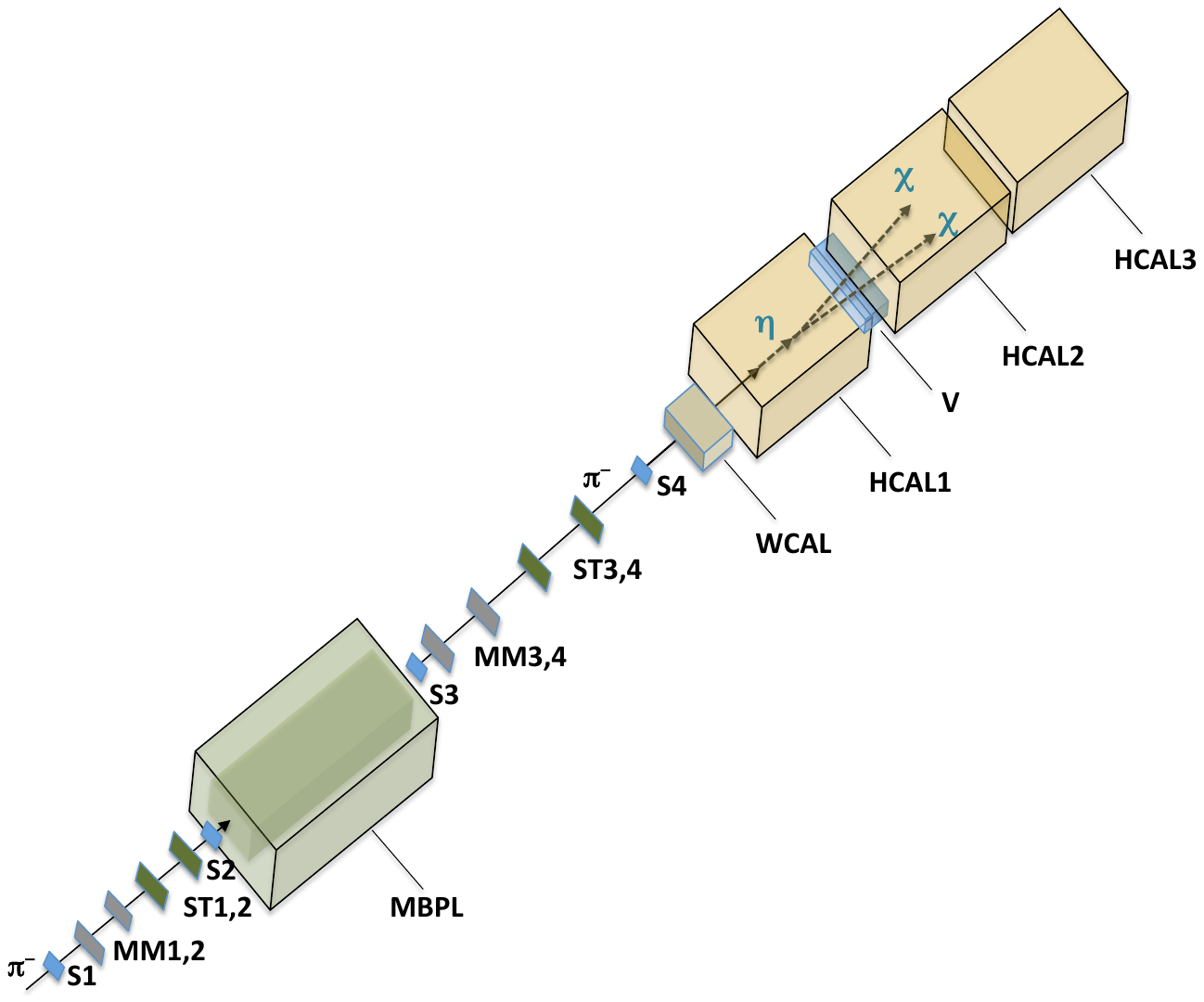}
\vskip-0.cm{\caption{Schematic illustration of the NA64h setup and of signal like event.\label{fig:setup}}}
\end{figure} 
Here,  the neutral meson is emitted mainly in the forward direction with the beam momentum and the 
recoil nucleon/nuclei carrying away a small fraction of the beam energy.
The term "quasi-elastic reaction" as applied to the process \eqref{eq:etaprod} means that, unlike elastic reactions of charge exchange with the proton, the transition can occur for the target  nucleus as a whole into an excited  state followed by its fragmentation. Since the binding energy in the nucleus is a few MeV/nucleon, the velocity $v\sim q/$mass of the daughter particles, where $q\lesssim0.03$ GeV/c is the momentum transfer, is on average small. At high initial energies, the nucleus does not have time to collapse during the interaction (the characteristic transverse distances is $l\simeq 1 /q$). After the collision, the nucleus disintegrates into fragments, which are absorbed into the target. Hence the experimental signature of the reaction \eqref{eq:etaprod}, is an event with  {\it  full disappearance of the beam energy}. The  decay $\etainv$ is expected to be a very rare event that  occurs with a much smaller frequency than the  $\etap$ production rate. Hence, its observation presents a challenge for the design and performance of the detector. However,  despite a relatively  small $\etap$ production rate  the  signature of the signal event \eqref{eq:etaprod} is very powerful  allowing  a strong background rejection. 

\par  The schematic of the NA64h detector modified  for a sensitive search for the reaction \eqref{eq:etaprod} is shown in Fig.~\ref{fig:setup}. The experiment employs the H4 50 GeV  pion beam at the CERN SPS with intensity $\simeq 10^6$ $\pi^-$ per  SPS spill of 4.8~s~\cite{h4}. The  $K^-$ contamination in the beam at the production target
($K^-/\pi^- \simeq 5\times 10^{-2}$)  is  reduced to $\simeq 2.5\times 10^{-2}$ at the detector location due to $K^-$ decay in flight \cite{na64beam}. 
 The beam is defined by the scintillator (Sc)  counters $S_{1-4}$. 
A  magnetic spectrometer (MBPL) is used to reconstruct the momentum of the incoming  $\pi^-$'s with a precision $\delta p/p \simeq 1\%$ \cite{Banerjee:2015eno}. The spectrometer consists  of two  consecutive dipole magnets with a total magnetic field of $\simeq$7 T$\cdot$m  and a  low-material-budget tracker composed of a set of  two upstream Micromegas (MM$_{1,2}$)  and two straw-tube chambers (ST$_{1,2}$), and two downstream MM$_{3-4}$, ST$_{3,4}$. 
 Downstream, the setup  is  equipped with  a  tungsten electromagnetic calorimeter (WCAL) of $\simeq 10$ radiation  lengths ($X_0$) to reject low-energy electrons in the beam. The hadronic calorimeter (HCAL) is composed by three modules (HCAL$1-3$),  
 for the  measurement of the full energy  $E_{HCAL} $ deposited by the beam. 
  Each HCAL module  
 is  a  matrix of $3\times 3$  cells assembled from  Fe and Sc plates  of $\simeq 7.5$ nuclear interaction lengths ($\lambda$).  The HCAL1 is followed by  a  high-efficiency veto counter~(V) to  enhance the rejection of events  with  hadronic secondaries produced in the $\pi^-$--nuclei interactions  in the target.  
 

 The key point of the design  is that the HCAL serves simultaneously as an active dump target composed mostly of 
 Fe(26,52) nuclei,  and a  massive, hermetic detector absorbing all the secondaries from the reaction  $\pi^- + Fe \to anything$, which occurs typically at the  first $\lambda$ of the HCAL1.  
 \begin{figure}[tbh!!]
\centering
\includegraphics[width=0.5\textwidth]{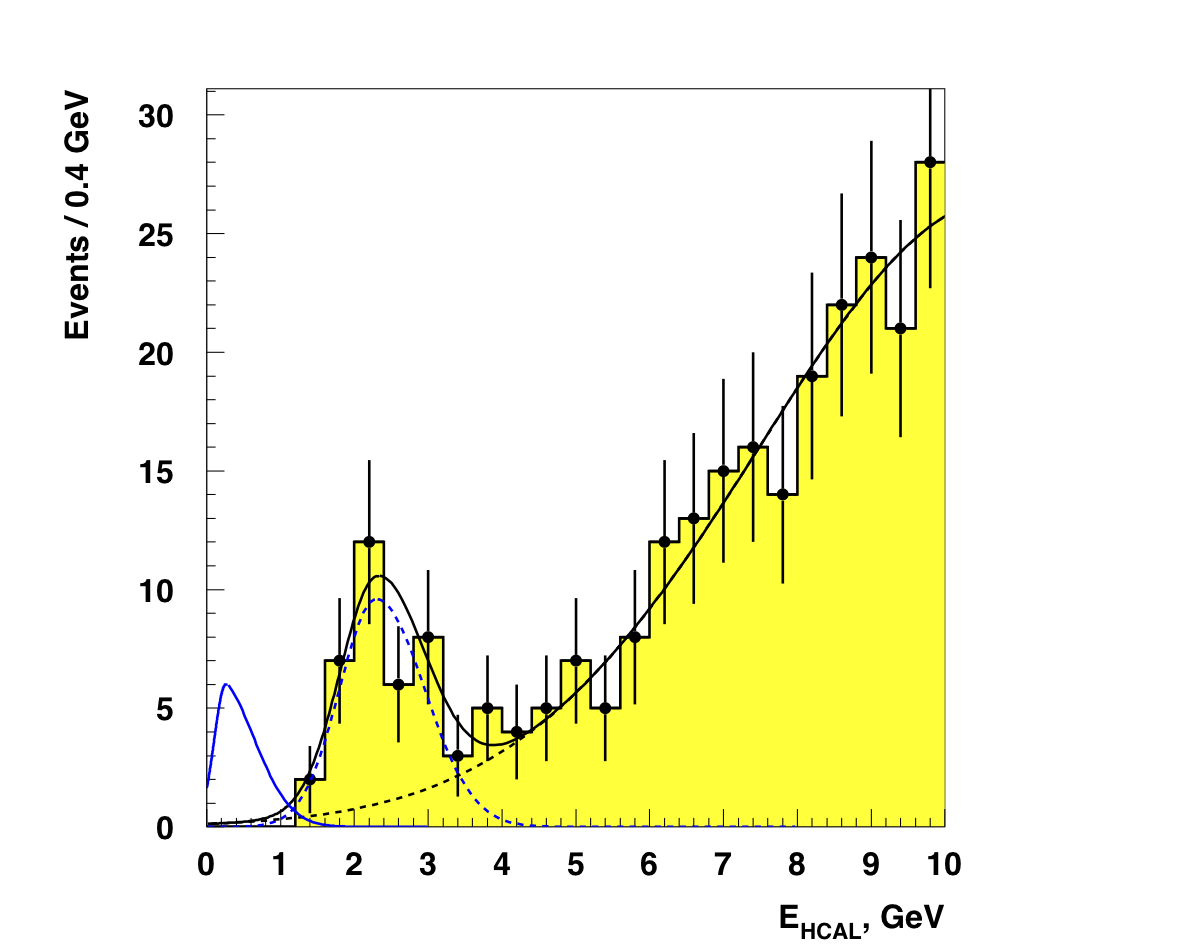}
\caption{The measured  distribution of the total energy deposited in three  HCAL modules (HCAL1+HCAL2+HCAL3) with signal selection cuts applied, i.e for events with the energy deposited in Veto $E_{VETO}\lesssim 2$ MeV.  The expected total background contributions (solid black) are overlaid on the data (points) from the shaded area. The peak at $\sim 2.5$ GeV  (dashed blue) corresponds to the background component due to  backward muons from $K^- \to \mu^- \bar\nu_\mu $ decays, the smooth  background above $\gtrsim 3$ GeV is the energy deposited in the HCAL by  poorly vetoed muons from  $K^- \to \mu^- \bar\nu_\mu $ decaying forward. The shape of the expected signal distribution from the $\eta \to invisible$ decay (solid blue) is normalized to ten signal events. For the $\eta' \to invisible$ the signal shape is very similar.}
\label{fig:signal-fit}
\end{figure}
\par Our data sample of $2.93\times 10^{9}$ pions on target was collected with the trigger requiring  the HCAL energy  $E_{HCAL} \lesssim 20$ GeV.  A  Geant4 \cite{Agostinelli:2002hh, geant} based Monte Carlo (MC) simulation package DMG4 \cite{dmg4,Oberhauser:2024ozf}
is used to study the performance of the detector,  define the selection cuts, estimate the signal acceptance and  background level. 
To maximize the signal/background ratio, the following   selection criteria  were used:
(i) The incoming track  must have momentum $50\pm 5$ GeV and the deflected track angle should be within 3 mrad  to reject events from the upstream $\pi^-$ interactions. 
(ii) There should be no multiple hits in the ST$_{3,4}$ chambers to reject events with charged secondaries  produced in the upstream beamline  material.
(iii) The  WCAL  energy should be within the range $\simeq 100 - 200$ MeV  expected from a  a minimum ionizing particle (MIP). 
(iv) The HCAL  energy deposit  should be $\lesssim 20$ GeV (trigger condition).
(v) The VETO energy must satisfy  $E_{VETO}  < 2 $ MeV ($\sim$30\% of the average MIP energy).
The low energy part of the HCAL energy distribution from  $\simeq 9.6 \times 10^5 $ events satisfying  these criteria,  is shown in Fig.~\ref{fig:signal-fit}. 
 \begin{table}[tbh!] 
\begin{center}
\caption{Expected background for $2.93\times 10^9$ pions on target.}\label{tab:bckg}
\vspace{0.15cm}
\begin{tabular}{lr}
\hline
\hline
Background source& Background, $n_b$\\
\hline
(i) Backward muons from $K_{2\mu}$ decay  &$ 0.08\pm 0.03$\\
(ii) Energy loss from primary pion  interactions  &$0.27\pm 0.1$\\
(iii) Punch-through leading $n,K^0_L $ & $<0.01$\\
\hline 
Total $n_b$    &    $0.35\pm 0.13$\\
\hline
\hline 
\end{tabular}
\end{center}
\end{table}
\par The signal  events from the reaction \eqref{eq:etaprod} are expected to deposit no or small energy in the HCAL  target.  
The latter can result from:
(i)~the HCAL noise and pileup events, measured directly with the random trigger; (ii)~ionisation losses of primary pion inside the target prior the pion interaction, 
defined from simulations and cross-checked with the energy deposited  by punchthrough muons; and (iii)~poorly reconstructed recoil neutrons or nuclear fragments resulting from the reaction \eqref{eq:etaprod}. 
This contribution was  simulated by using the measured differential cross sections $d\sigma/dt$ of the reaction  \eqref{eq:etaprod} as a function of the momentum transfer $t$, assuming that the deposit  from the nucleus break-up is small \cite{apok87, sng2023}. 
Finally,  all the (i) - (iii) contributions were summed up defining  the shape of the signal from the reaction \eqref{eq:etaprod}, 
 which is shown in Fig.~\ref{fig:signal-fit}.  
\par The total background in the signal region $\lesssim 1$ GeV  has several  components:
(i) The main  and most sophisticated  background,  peaking around $\sim$ 2.5 GeV as  shown in Fig.\ref{fig:signal-fit},  is from the  $K^- \to \mu^- \bar\nu_\mu$  decay in flight after the magnet with the  muon momentum directed backward, in the $K^-$ rest frame, with respect to the  beam momentum. The momentum of these muons is constrained by the geometrical acceptance of the S4 counter in the  $2.2 < P_mu < 3.4$ GeV  range, see Fig.\ref{fig:setup} and most of them stop in the HCAL1 resulting in a broad peak as 
shown in Fig.\ref{fig:signal-fit}.   These low-energy decay 
muons penetrate the WCAL depositing energy via ionization and stop in the HCAL1 module with the  energy deposition close to the signal region.
Those muons from  $K^- \to \mu^- \bar\nu_\mu $  decay that decays perpendicular to the $K^-$ direction, typically escape detection as they do not trigger the S4 counter. 
 The number of   $K^- \to \mu^- \bar\nu_\mu $  decays is  estimated from the simulations considering the beam composition measurements from \cite{na64beam}.
(ii) The second bulk component includes events from  the HCAL low-energy tail  $E_{HCAL} \lesssim 10$ GeV which deposited smaller energy due to the energy leak 
resulting from secondary $\pi,K \to e (\mu)+ \bar\nu_e (\bar\nu_\mu)$ decays in the target and  escaping neutrals. A small contribution also arises from  
 $K^- \to \mu^- \bar\nu_\mu$ decays when the muon momentum in the rest frame of the kaon is in the forward direction. These muons can trigger the S4 counter with a momentum typically  $P_\mu \gtrsim 10$ GeV with a lower energy tail due to the beam  divergence and  poor energy reconstruction.
    The Veto does not detect a small fraction of these muons ($\lesssim 10^{-3}$). This inefficiency  was estimated from the measurements with the single muon punch-through in the HCAL.
     (iii)    Punch-through of leading neutral hadrons $(n, K^0_L)$ from the $\pi^-$  interactions in the target estimated from the direct measurements of punch-through events 
     as described in Ref.\cite{na64alp} is negligible.
 After applying the  cuts, we expect mostly background events of type (i) and (ii)  to remain in the  analysis. 

\par To obtain  upper limits on the $\eta,\eta' \to invisible$ branching ratios,  the analysis of events  using the technique of  limit setting
 based on the RooStats package \cite{root} was performed. 
 First, to define the optimal signal box, the expected signal shape and background level, efficiencies,  and uncertainties were used for comparing sensitivities calculated as a function of  the HCAL energy cut $E_{cut}$. The sensitivity was   defined as an average expected limit calculated using the profile likelihood ratio method.
 To reduce uncertainties and the dependence on MC simulations, the  expected number of background events $n_b$  in the signal box was finally obtained directly from 
  the data.  The measured HCAL energy distribution was fitted to the sum of two functions  $f_0 = f_1 + f_2$, where $f_1$ describing  the backward decay muon peak and $f_2$  the  second bulk event component, respectively, as depicted in Fig. \ref{fig:signal-fit}. The shape of the extrapolation functions was taken from the analysis of the data and cross-checked with simulations by normalizing the peaking background component in the MC  to the number of data events. 
  The  evaluation of $n_b$  was  obtained  by extrapolating the function $f_0$ to the signal region. The systematic errors arising from the normalization and signal efficiency uncertainties were assessed by varying the fit functions.  
Finally, the candidate events were requested  to have HCAL energy  $E_{HCAL}  \lesssim  0.85$ GeV, thus, for the background extrapolation only energy deposition greater than this threshold  in the active target is considered.   The estimated background inside the signal region was  0.35$\pm$0.13 events.   After determining all the selection criteria and background levels, no event is found in the signal region. 


\par  The upper limits for  ${\rm Br}(\etap \to invisible)$ are  obtained 
from the 90\% C.L.  upper limit for the expected number of signal events, $n_{\etap}^{90\%}$ 
 by applying the modified frequentist approach for confidence levels, considering the profile
likelihood ratio as a test statistic \cite{junk,limit,Read:2002hq}. The number  of signal events $n_{\eta^{(\prime)}}$ from the $\eta^{(\prime)} \to invisible$ decay in the signal box is given by:
\begin{equation}
n_{\etap} = n_{\pi} \epsilon_{tr} \epsilon_{\pi} \epsilon_{\etap} \frac{\sigma(\etap)}{\sigma_\pi(tot)}{\rm Br}(\etap \to invisible)
\label{eq:nev}
\end{equation}
where $ n_{\pi} = (2.93\pm0.04)\times 10^9$ is the number of incoming pions on target, $\epsilon_{tr} = 0.98\pm 0.02$,  $\epsilon_{\pi} = 0.47\pm0.023$, and  
$\epsilon_{\eta} =0.75\pm0.03$ and $\epsilon_{\eta'} =0.73\pm0.03$  are respectively the trigger efficiency, selection efficiency for the incoming  pion, and the signal efficiencies for the given energy cut $E^{cut}_{HCAL}$.
The cross sections $\sigma(\pi^-,\etap)$ of the reaction \eqref{eq:etaprod} were evaluated from the set of direct measurements of the charge exchange  reactions on H, Li, C, Al, and Cu nuclei for the $\pi^0$, $\eta$, and $\eta'$ final  states  for the   beam energies up to 50 GeV 
\cite{bolot74-1,bolot74-2,bolot75,apeleta,apeletapr,flam,daum, yud1, apok82,apok87}  and 
found to be  $\sigma(\pi^-,\eta) = 21.9\pm  7.5~\mu$b and  $\sigma(\pi^-,\eta') = 10.4\pm 3.5~\mu$b for Fe nuclei at 50 GeV as described in \cite{sng2023}. The total cross-section for the 50 GeV pion 
absorption on Fe target, $\sigma_\pi(tot) = 554 \pm 16$ mb was evaluated from the measurements performed for  different target nuclei and a wide range 
of energies in Ref.\cite{carroll}.  
The signal acceptance was calculated by taking into account  the  shape of the signal distribution in the target. 
Finally, the 90\% C.L. exclusion limits on $\eta, \eta' \to invisible$ decays,  
\begin{eqnarray}
&{\rm Br}&(\eta \to invisible) <1.1 \times 10^{-4}  \label{eq:breta} \\
   &{\rm Br}&(\eta' \to invisible) < 2.1 \times 10^{-4}, \label{eq:breta'}   
\end{eqnarray}
obtained by taking into account the estimated background and systematic errors for the efficiencies in Eq.(\ref{eq:nev}) are dominated by the  uncertainty of $\sim 34 \%$  in
  the $\eta$ and $\eta'$ production cross sections. 
\par  In summary, the  proof-of-concept search with NA64h places the first constraints on $\eta, \eta' \to invisible$ decays using charge exchange reaction as a source of $\eta$ and $\eta'$ mesons and missing energy  as a powerful signal signature. Our limit  of 
Eq.(\ref{eq:breta}) is comparable, while the limit of Eq.(\ref{eq:breta'}) is more stringent  by  a factor of $\simeq3$ compared to the current best limits set by BESIII  \cite{besiii}.
 These results demonstrate the effectiveness of our approach.  
 Improving the beam  quality by installing Cherenkov counters to suppress the  kaon component, using a high-granularity active  target, 
extending running times,  and enhancing the background characterization are all concrete avenues to further improve the  sensitivity in future searches. A significant improvement  in the accuracy of the measurement of the cross section \eqref{eq:etaprod}, compared to our current knowledge  \cite{sng2023} would also 
reduce significantly the main systematic source of the measurement.  Finally, our method could  also be used to search for leptophobic dark sectors in invisible decays of vector mesons \cite{toro,aszh}. 

\par  We gratefully acknowledge the contributions to this experiment by M. Kirsanov, discussions  with V. Valuev,  
the support of the CERN management and staff, as well as  contributions from  
HISKP, University of Bonn (Germany), ETH Zurich and SNSF Grant No. 186181, No. 186158, No. 197346, No. 216602 (Switzerland),  
ANID - Millennium Science Initiative Program - ICN2019 044 and Grants FONDECYT No. 1240216, No. 1240066, No. 3230806 (Chile),  RyC-030551-I and PID2021-123955NA-100 and CNS2022-135850 funded by MCIN/AEI/ 10.13039/501100011033/FEDER, UE (Spain). 
This result is part of a project that has received funding from the European Research Council (ERC) under the European Union's Horizon 2020 research and innovation programme, Grant agreement No. 947715 (POKER).

\end{document}